\documentclass[12pt]{article}
 \usepackage{amsfonts}
 \usepackage{amssymb,amsmath}
 \usepackage{graphicx} \usepackage{epstopdf}
 \newcommand{\crlb}[1]{\label{#1}\\[2pt]}
 
 \newcommand{\crld}[1]{\label{#1}}
 \newcommand{\eela}[1]{\quad\hbox{\scriptsize{#1}}\label{#1}\end{eqnarray}}
 \newcommand{\eelb}[1]{\label{#1}\end{eqnarray}}
 
 \newcommand{\newsecb}[2]{\section{#1}\label{#2}\setcounter{equation}{0}}

 \newcommand{\nolabels} {\def\eel{\eelb}\def\eeql{\eeqlb}  \def\crl{\crlb} 
 \def\newsecl{\newsecb}\def\bibiteml{\bibitem} \def\citel{\cite}\def\labell{\crld}}
\newcommand{\eeqla}[1]{\quad\hbox{\scriptsize{#1}}\label{#1}\end{aligned}\end{equation}}
\newcommand{\eeqlb}[1]{\label{#1}\end{aligned}\end{equation}}

\newcommand\publishversion  {\nolabels\setlength{\textheight}{8.38in}\setlength
    {\oddsidemargin}{0in} \setlength{\textwidth}{6.2in}\setlength{\topmargin}{-0.2in}}

\def\beq{\begin{equation}\begin{aligned}}		\def\eeq{\end{aligned}\end{equation}}
\def\be{\begin{eqnarray}}  					\def\ee{\end{eqnarray}}		
   \def\bi#1{\begin{itemize}\item[#1]} 	     \def\itm#1{\item[#1]} 	   \def\ei{\end{itemize}} 
   \def\eqn#1{(\ref{#1})}
\def\Tr{{\mbox{Tr}}\,}   	 \def\fn{\footnote}	 

		 \def\del{\delta}  

 \def\bet{\beta}        
 
 \def\del{\delta}      \def\Del{\Delta}    
             \def\m{\mu}
             \def\vv{\varphi}    
             \def\rr{\varrho}       
      \def\tht{\theta}  
               
       \def\W{\Omega}

     \def\OO{{\mathcal O}}  
     
 \def\pa{\partial} \def\ra{\rightarrow} 
 \def\bal{$\bullet$} 
 \def\dd{{\rm d}}  \def\bra{\langle}   \def\ket{\rangle}

\def\bal{$\bullet$} 
\def\fract#1#2{{\textstyle\frac{#1}{#2}}}	 	 	
\def\ffract#1#2{\raise .2 em\hbox{$\scriptstyle#1\,$}\kern-.34 em/\kern-.34 em\lower .15 em \hbox{$\scriptstyle\,#2$}}
\def\half{\fract12}					
\def\tl#1{\tilde{#1}} 
\def\ex#1{e^{\textstyle#1}} 			
\def\bpmatrix{\begin{pmatrix}} 			\def\epmatrix{\end{pmatrix}}
\def\bmatrix{\begin{matrix}} 			\def\ematrix{\end{matrix}} 
\def\bcenter{\begin{center}}			\def\ecenter{\end{center}}

\def\lowerheightgth#1#2#3{\(\raise-#1\hbox{\includegraphics[height=#2]{#3}}\)}
\def\lowerwidthgth#1#2#3{\(\raise-#1\hbox{\includegraphics[width=#2]{#3}}\)}
\def\widthgth#1#2{\includegraphics[width=#1]{#2}}
\def\th{\({}^{\mathrm{th}}\,\)}

\def\inn{{\mathrm{in}}} \def\outt{{\mathrm{out}}} 
 \def\Hawk{{\mathrm{Hawking}}}
  \def\BH{{\mathrm{BH}}}

\def\weglaten#1{}	
 
  \publishversion
\begin{document}
\begin{titlepage} 
\title{
How studying black hole theory\\  may help us to quantise gravity\fn{presented at the Conference on ``Eternity between and Space and Time'',  Padova, May 19-21, 2022.}
\author{Gerard 't~Hooft}}
\date{\normalsize
Faculty of Science,
Department of Physics\\
Institute for Theoretical Physics\\
Princetonplein 5,
3584 CC Utrecht \\
\underline{The Netherlands} 
http://www.staff.science.uu.nl/\~{}hooft101}
 \maketitle

\begin{quotation} \noindent {\large\bf Abstract } \\[10pt]
 Black holes are more than just odd-looking curiosities in gravity theory. They uniquely intertwine the basic principles of General Relativity with those of Quantum Theory. Just by demanding that they nevertheless obey acceptable laws of dynamics, just like stars and planets, we hit upon strange structures that must play key roles in the quantum effects that we expect in the gravitational force at ultrashort distance scales.\\
 It is explained why, in our approach to address the problem of information conservation, 
  the usual expression for the temperature of Hawking's radiation is off by a factor 2.
 
 \end{quotation}\end{titlepage}

\newsecl{Introduction}{intro} \setcounter{page}{2}
Gravitation is arguably the most elementary force in physics. It appears 
		to be directly linked to a fundamental principle:
	\begin{quote}Invariance of physical laws under general 
		coordinate transformations in space and time.\end{quote}
As is well-known, other forces among the elementary particles of matter are based on very similar symmetry principles, notably those of local gauge theories, but the symmetry principle that governs the gravitational force seems to be at the basis of all symmetries in nature, that of general transformations among all coordinates for space and time. And more remarkably,
	Combining this symmetry principle with 
	 quantum mechanics seems to lead to novel and unique clashes with what we thought we knew about forces and dynamics.
	 
	 It was thought that history of science had given us a significant clue: \emph{``We will get this thing done within half a century or so !}
	 
	 Unfortunately, in spite of a tremendous amount of work and numerous essential, novel pieces of insight, we still are confronted with mysteries at this point. What are we doing wrong?

	Most conventional theories, focussing either on physics at the Planck scale (some \(10^{-33}\) cm or \(10^{-44}\) seconds, or on the scale of cosmological theories (ranging up to the size of the universe), assume that the quantum formalism is mandatory: 

	\emph{Start with the existence of a Hilbert space}, formulate a law for computing quantum amplitudes, and assume that all physical phenomena covered by the theory can be described in terms of these amplitudes, \emph{ even if } there is no need to agree on what it is that these amplitudes actually describe. According to many researchers, discussions on the foundations of quantum mechanics itself have come to a dead end. Here however, we emphasise that quantum amplitudes can actually be reduced to being a \emph{vector representation} of phenomena based on extremely mundane forms of logic.\,\cite{GtHCA}
	 One may ignore the usual conundrums of the `collapse of the wave function', the role played by `pilot waves', and even the existence of uncountably many distinct universes (the `many world' hypothesis).
	
	And indeed we must worry about the nature of general coordinate transformations that we thought to have under control.
		
A deeper study of black holes\,\cite{Schwarzsch.ref}, and their relation with the laws of quantum mechanics\,\cite{Hawking.ref}, may teach us new and very important things. Black holes seem to be just the most basic solutions of Einstein's field equations, but it is not automatically guaranteed that their connection with quantum mechanics will be anything ordinary. In our telescopes it seems that black holes may be just a special kind of burnt-out stars. 
But if conventional theories for the quantum treatment are not completely wrong, one does not get `ordinary behaviour' from our equations. Something is wrong. So, we advertise a closer look at those laws. Just by postulating that everything hangs together in a very orderly manner, may force us to rephrase those equations, in a way that might become very revealing.	

\newsecl{Schwarzschild and the tortoise metric}{BHmetric.sec} 
Much of the material described in this section and the next one has appeared in several previous reports by this author.\cite{GtHQuStr.ref,GtHStringsgrav.ref,GtHclones.ref}\  Readers familiar with is can just briefly scan this part of the paper, but we shall refer to it when we continue.

Shortly after Einstein published his theory of General Relativity, the astronomer Karl Schwarzschild\,\cite{Schwarzsch.ref} realised that assuming spherical symmetry enables us to write down an exact solution of these equations. The metric one arrives at, in modern notation, reads
	\be \dd s^2=-\dd t^2\Big(1-\frac{2GM}{r}\Big)+\frac{\dd r^2}{1-\frac{2GM}{r}}+r^2\dd\W^2\ , \eel{Schw.eq}
	where \(r\) represents distance from the origin, \(t\) is a time coordinate, and \(\W\) stands for the 
	solid angle coordinates \((\tht,\vv)\). The variables \(\dd r\) and \(\dd t\) stand for infinitesimal differences for the coordinates of two adjacent points in space-time, and \(\dd\W\) stands short for their infinitesimal angular 
	separation, \(\dd\W^2=\dd\tht^2+\sin^2\tht\,\dd\vv^2\). \(M\) is a parameter that stands for the total mass
	as would be perceived by an observer at infinity. See Fig.~\ref{Schw.fig}a.

	\begin{figure}[h!] \begin{center} 
		\vskip 20pt time \(\uparrow\) \hskip-10pt 
		\lowerwidthgth{90pt}{150pt}{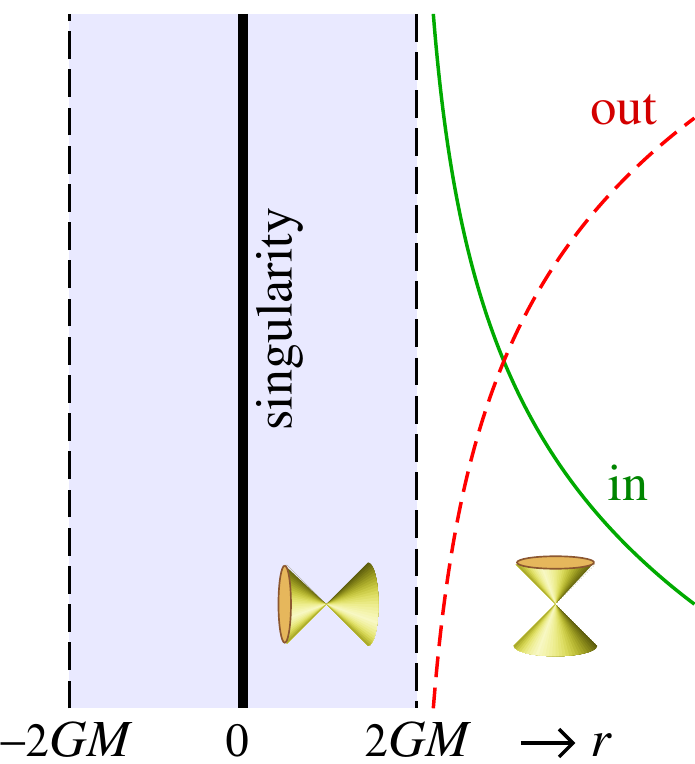} \qquad \quad\lowerwidthgth{80pt}
		{150pt}{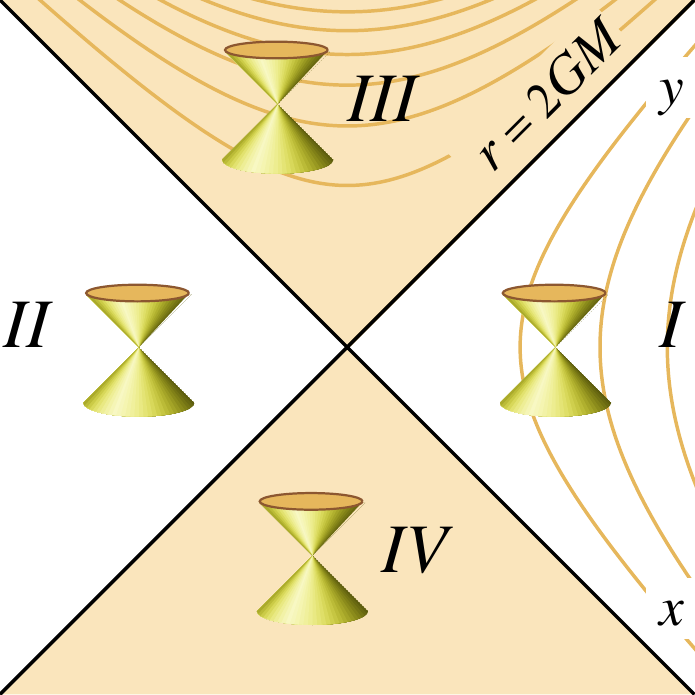}  \\
		\hskip-100pt a) \hskip 230 pt b)\end{center} \vskip-20pt
		\begin{caption} {a:	The Schwarzschild metric. Vertical dashed lines: the event horizon. 
		Curved lines: light like radial geodesics: \(y\,=\,\)const (smooth), and \(x\,=\,\)const (dashed). 
		Cones: the orientation of the local light cones. 
		Angular coordinates \((\tht,\,\vv)\) are not shown. \quad
		b: The Tortoise coordinates \((x,\,y)\), showing regions \(I - IV\) and the orientations of the
		 local light cones. Curves show \(r=\)\,const. lines.}\labell{Schw.fig} 
		 \end{caption}
	\end{figure}

This metric is independent of time \(t\), and we see that the contributions from \(\dd r\) and \(\dd t\), as given in the first two terms in \eqn{Schw.eq} both switch signs when \(r\) crosses the point \(r=2GM\). 
Here, G is Newton's gravitational constant, and the freely adjustable parameter \(M\) represents the mass of this object. Much later, it would become known as a `black hole'.
	
The apparent singularity at \(r=2GM\) is not real, it is a coordinate artefact. One can replace the longitudinal coordinates \(r\) and \(t\) by a new set, \(x\) and \(y\), see Fig.~\ref{Schw.fig}b, as follows:
	\be x\,y&=&\Big(\frac r{2GM}-1\Big)\,\ex{r/2GM}\ ; \\[4pt]
	y/x&=&\ex{t/2GM}\ . \eel{tortoise.eq}
These coordinates are called after Kruskal\,\cite{Kruskal.ref} and Szekeres\,\cite{Szekeres.ref}, and shall be referred to here as  `tortoise coordinates' for short, just because they replace the infinitely slow geodesics to and from the horizon, by ordinary geodesics crossing a light front.

In these coordinates, the singularities at \(r\ra 2GM\) disappear:
	\be \dd s^2\ =\  \frac{32(GM)^3}r\,\ex{-r/2GM}\dd x\,\dd y\ +\ r^2\dd\W^2\ .  \eel{KSmetric.eq}
However, here we discover that there are two horizons, not one: the \emph{future event horizon}, at \(x=0\), to which all absorbed matter particles move, and the \emph{past event horizon}, at \(y=0\), which may emit particles.
The tortoise coordinates show that the Schwarzschild space-time has a natural extension, from region \(I\),
where \(x>0\) and \(y>0\), to regions \(II,\ III,\) and \(IV\). Region \(I\) is the \emph{physical region}. Here, we can draw the geodesics for all in-going particles, and all out-going particles, see Fig.~\ref{regionI.fig}.

\begin{figure}[h] \begin{center} \vskip-10pt
		\widthgth{250pt}{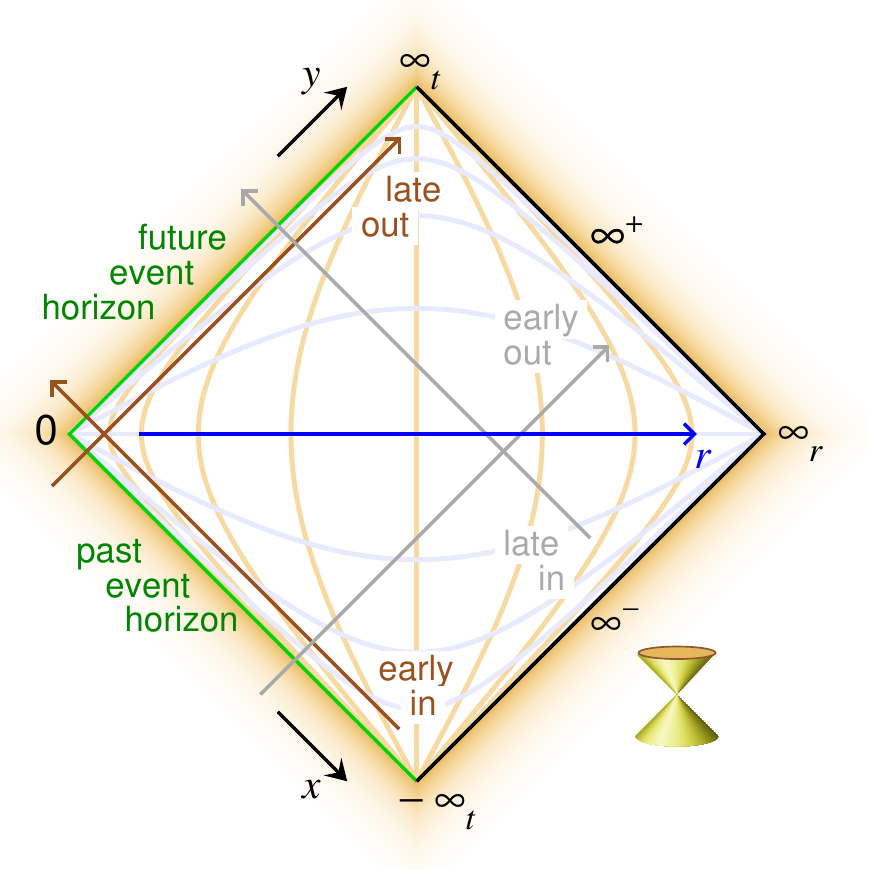} \end{center} \vskip-20pt
		\begin{caption} {Region \(I\), the physical region (see Fig.~\ref{Schw.fig}b), showing early and late particles going in and out.
		Everywhere in this region the light cones are oriented as the one shown.}
		\labell{regionI.fig}\end{caption}
	\end{figure}
	
Believing that black holes should \emph{not} be approached as magic wands, but should be understood as normal physical objects that can absorb things and emit things, leads one to believe that in-particles (short for particles going into the black hole) transmit all information they contain towards the out-particles. At first sight, this seems to violate causality, since the points where in-particles cross the future event horizon lie in the far future while the out-particles emerge in the far past  (see Fig.~\ref{regionI.fig}).

However, they meet each other half way, and if this point lies much further away than a Planck length from the horizon crossing point (point 0 in the figure), then the contradiction becomes insignificant. This separation decreases exponentially with the time span lying between the in-orbit and the out-orbit, so the Planck domain is reached quite quickly. But more to the point, we can compute what happens here. It is the \emph{gravitational force between in-particles and out-particles that does the job of information transfer}  (as soon as their longitudinal coordinates meet), and, it acts locally as a sonic boom; there is no contradiction with causality at all.

\newsecl{The Shapiro effect}{Shapiro.sec}
How, by gravity, an in-particle  1 affects an out-particle  2 -- and vice versa -- is an elementary exercise in General Relativity. Consider first a flat background. One can take  particle 1 to be at rest, and consider its metric as a Schwarzschild metric. Then let this particle move with a velocity close to \(c\). Suppose now another light particle, 2, moves in the opposite direction. It experiences the gravitational field of particle 1 as being highly compressed by the Lorentz transformation. Immediately before, and immediately after the encounter, space-time is experienced as being flat (in the home frame of particle 1,  the particles are then far apart). So the gravitational field of particle 1 is experienced as a sonic boom, comparable to Cherenkov radiation.\cite{Cherenkov.ref}

During this encounter, what happens is closely related to the Shapiro effect\,\cite{Shapiro.ref} when a light ray grazes past the Sun or another heavy body. The geodesic of the light ray is delayed. An elementary calculation\,\cite{AichelbSexl.ref} shows that this amounts to a shift \(\del u^\m\) of   particle 2 in the light cone direction of the motion  of particle 1. If particle 1 has a light cone momentum \(p^\m\) then one finds
	\be \del u^\m=-4G\, p^\m\,\log|\tl x-\tl x'|\ . \eel{Shshift.eq}
Here, \(\tl x-\tl x'\) is the transverse separation at the closest encounter (impact parameter).

This was the effect in a flat background space-time. Now consider particles moving in and out of a black hole\,\cite{GtHDray.ref}, entirely in the longitudinal direction (close to the horizons, the transverse motion, and shift, are neglected for the time being).  For a particle 1 moving in the longitudinal \(p^-\) direction, this shift can be written as\,\cite{GtHQuStr.ref,GtHStringsgrav.ref}
	\be	\del u^-(\W)=8\pi G\,f(\W,\W')\,, \eel{BHshift.eq}
where \(\del u^-\) is the shift in the orbit of particle 2. The function \(f\) replaces the logarithm in Eq. \eqn{Shshift.eq}. It is a Green function obeying
	\be (1-\Del_\W)f(\W,\W')=\del^2(\W,\,\W')\ . \eel{fGreen.eq}
An in-particle with momentum \ \(p^-\)  \ at solid angle \ \(\W'\) causes a shift \(\del u^-\) at solid angle \(\W\)\,.	
This Shapiro effect has a very important property: \emph{it is linear}: the effects of all in-particles is just the sum of the contributions of every single one. At the same time, it displaces all out-particles by the same position-dependent amount \(\del u^-(\W)\).

This enables us to write the combined effect as a property of functions on the angles \(\W,\ \W'\).	
Write a single function   \(p^-(\W)\) as
	\be p^-(\W)\equiv \sum_i p^-_i\del^2(\W,\W_i)\ ,\eel{pminW.eq}
where \(\W_i\) are the spots where the in-particles enter the future event horizon; for all out-particles we write:
	\be u^-(\W_i)=u^-_i\ . \eel{uminW.eq}
The Shapiro effect can now be written as follows: if we add an in-particle with momentum \(\del p_i^-\) at solid angle \(\W_i\) then it modifies the position \(u^-(\W)\) of all out-particles, in accordance with the equations
	\be u^-(\W) = 8\pi G\int\dd^2\W')\,f(\W,\W')\,p^-(\W')\ ,\eel{dufunct.eq}
where the Green function \(f(\W,\W')\) obeys \eqn{fGreen.eq}.
We can write
	\be (1-\Del_\W)u^-(\W)=8\pi G\,p^-(\W)\ . \eel{Deluminus.eq}
Notice that we replaced the shift equation \eqn{BHshift.eq} for one particle going in and one particle going out, by a single functional equation \eqn{dufunct.eq} for all particles in and all particles out. The only difference this makes is that we could omit the symbol \(\del\) that indicates \emph{change}. Before and after any change, out-positions relate to in-momenta in the same way. The equation suggests that the origin, where all positions are zero and no particles are going in at all, should be chosen as the origin of this space.

\newsecl{Quantum mechanics}{QM.sec}
	In the previous section, we encountered fundamental new equations relating positions and momenta of in and out going particles. This begs for an investigation as to what happens when we introduce quantum mechanics. Close to the origin\fn{The process of information transfer, in which we are now interested,  takes place at Planckian distance scales. As long as the Schwarzschild metric parameters, typically in the order of \(GM\), are large compared to this, our analysis applies.}, all particles obey
	\be {}[u^\pm_i,\,p^\mp_j]=i\del_{ij}\ ,\qquad [u^\pm,\,p^\pm]=0\ , \eel{qucomm.eq}
so that, for the functions \(u^\pm(\W)\) and \(p^\pm(\W)\),  we can write
	\be{}[u^\pm(\W),\,p^\mp(\W')]&=&i\del^2(\W,\,\W')\ ,\\[3pt]
	{} [u^\pm(\W),\,p^\pm(\W')]&=&0\ .\eel{functcomm.eq}

In combination with Eqs.~\eqn{dufunct.eq} and \eqn{Deluminus.eq}, these equations become very powerful. The equations are local now, and, most important, they are linear. Linearity and locality might get broken when we refine these equations for large transverse momenta, but we believe that the section for sufficiently low transverse momenta can be kept by itself as the dominant contribution to the information processing mechanism. 
Due to linearity, it is meaningful to expand the functions \(u^\pm\) and \(p^\pm\) in spherical harmonics:
	\be u^\pm(\W)\equiv\sum_{\ell,m}u^\pm_{\ell m}Y_{\ell m}(\W)\ ,\qquad p^\pm(\W)\equiv
	\sum_{\ell,m}p^\pm_{\ell m}Y_{\ell m}(\W)\,, \eel {harm.eq}
where the sum goes over all nonnegative integer values of \(\ell\), and \(-\ell\le m\le \ell\ .\)
	
	For the spherical harmonics, the operator \(\Del\) diagonalises into \(-\ell(\ell+1)\), so that we can write
	\be u^-_{\outt,\,\ell m}=\frac{8\pi G}{\ell^2+\ell+1}\,p^-_{\inn,\ell m}\,,\qquad 
	u^+_{\inn,\,\ell m} =-\frac{8\pi G}{\ell^2+\ell+1}\,p^+_{\outt,\ell m}\,,\eel{diagonal.eq}
where we wrote the subscripts `in'  and `out', to indicate that these operator refer to in- or out-particles.
The minus sign in the last equation is understood as a consequence of the antisymmetry of the commutator while no minus signs had been inserted in Eqs~\eqn{qucomm.eq}:
	\be [u^+_{\ell m},\,u^-_{\ell'\,m'}]=\frac{8\pi  i G}{\ell^2+\ell+1}\,\del_{\ell \ell'}\del_{mm'}\ . \eel{commuu.eq}
	
Important remarks: \bi\bal There is only one real independent dynamical variable \(u^+\), and a variable \(p^-\) canonically associated to it, at every value of the pair of integers \((\ell,\,m)\). So the variables in Eq.~\eqn{qucomm.eq}  are replaced by a one-dimensional quantum mechanical pair at each  \(\ell\) and \(m\).
\itm{\bal} Different \((\ell,\,m)\) values all commute. Therefore, in the approximations used, the complete set of all quantum states will be just the product of all quantum states \(|u^+\ket_{\ell m}\) at all \((\ell,\,m)\). By taking their one-dimensional Fourier transforms, one gets the states \(|p^-\ket_{\ell m}\) instead. \ei
Thus, the out-particles emerge as being the Fourier transforms of the in-particles, a very simple algebra at this stage.

Note that, replacing one single species of matter by a set of multiple matter species, would lead to contradictions (there exists only one type of gravitational force).
There exists only one form of matter which we subject to the rules \eqn{qucomm.eq}, and therefore we cannot apply second quantisation in the longitudinal direction; only the \(\W\) dependence can be viewed as a reduced, two-dimensional second quantisation process. Having single particle states only for the \(u^\pm\) dependence, but second quantisation only in the \(\tht,\vv\) direction could be called `1.5\th quantisation'.

\begin{figure}[h]\begin{center}
		\widthgth{400pt}{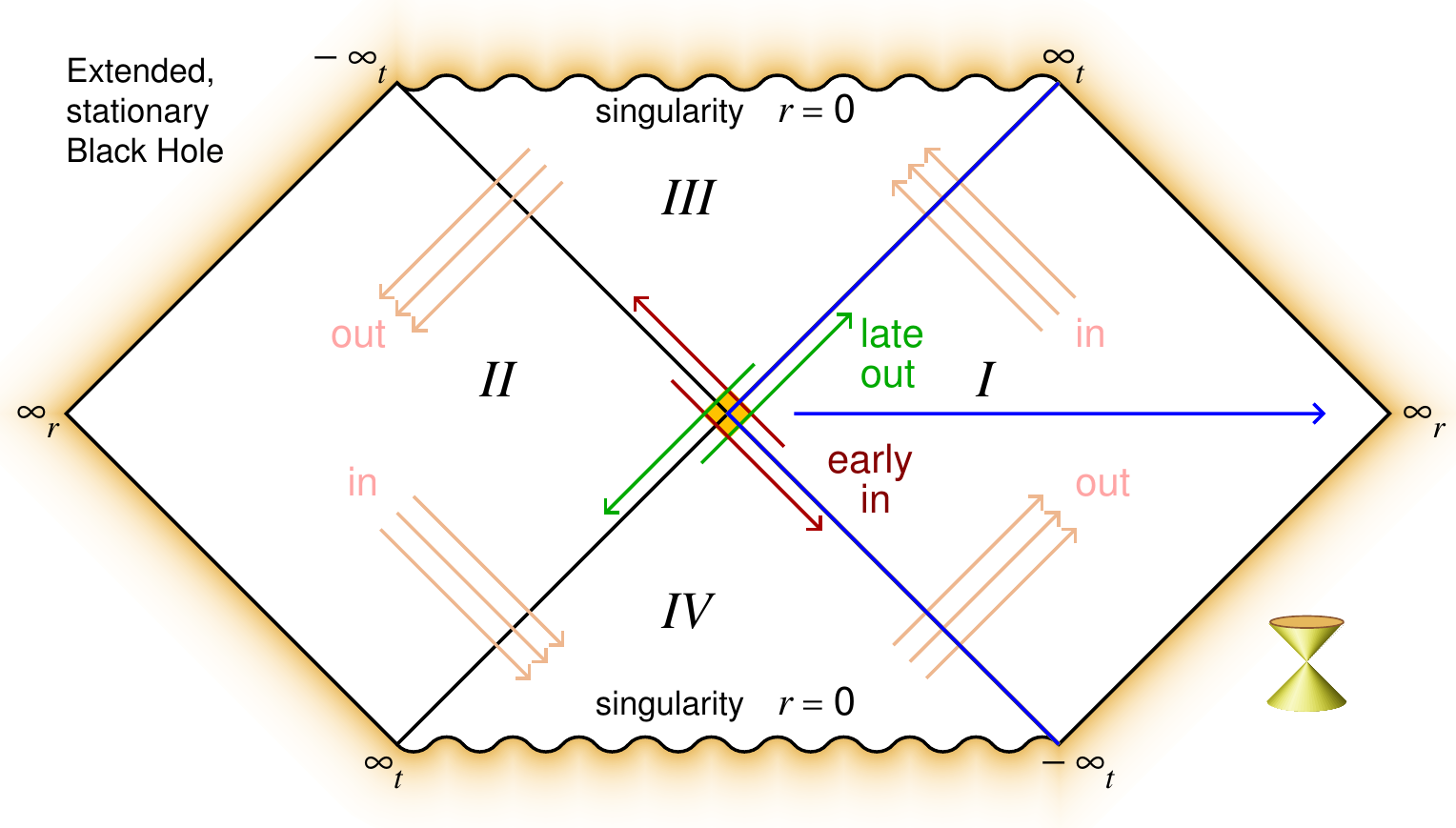} \vskip-23pt \end{center}
		\begin{caption}{Penrose diagram generated from the tortoise coordinates \(x\) and \(y\), by 
		mapping these both on a compact segment \([-1,1]\). See text.\labell{Penrose19.fig}}
		\end{caption}
	\end{figure}

Everything seems to come together, but there are still some hurdles to take. By mapping the Kruskal-Szekeres coordinates onto compact segments, one gets the Penrose diagram for an eternal black hole,
see Fig.~\ref{Penrose19.fig}. It was shown above that the out-particles are obtained from the in-particles
by Fourier transforming the wave functions, call these \(\bra u^+|\psi\ket_{\ell,m}\)\,. The Fourier transformation is a unitary process, so one might think that we obtained the complete black hole evolution operator \(U(t)\). The fact that, so-far, we concentrated only on \emph{eternal} black holes, is not a problem. At time scales longer than \(\OO(M\log M)\) in Planck units, the wave functions get completely scrambled by the Shapiro process, and the component of Hilbert space that would represent either collapse, or a final evaporation event, is expected to be small compared to the one spanned by the states that we do take into account.

A problem that we still have, at this stage, is that there are two asymptotic regions, \(I\) and \(II\). Physically, we need unitarity only for one single domain, region \(I\) itself. But by generating the shifts \(\del u\), the Shapiro process will transport states right across the horizon, and when we attempt to calculate the combined effects, we find that unitarity is violated by that proces: particle wave functions in region \(I\) will generate Fourier transforms that cover both regions \(I\) and \(II\).

It is true that, most of this process takes place in a Planckian domain close to the origin as indicated in Figure \ref{Penrose19.fig}, but, as time goes on, local Lorentz boosts will quickly send these out-states to infinity, spreading themselves both over regions \(I\) and \(II\).
	
In the past, we proposed to cure this problem by assuming region \(II\) to correspond to the \emph{antipodes} (in terms of the angles \(\tht\) and \(\vv\)) of region \(I\). But this did not work as assumed. In terms of the harmonic functions \(Y_{\ell m}(\W)\), going to the antipodes means multiplying the angular wave functions by \((-1)^\ell\). For odd \(\ell\), this actually cancels the minus sign transporting us from region \(I\) to region \(II\) or back, causing our calculations to fail.

We now believe that region \(II\) is an \emph{exact quantum copy}\,\cite{GtHclones.ref} of all states in region \(I\). This means that we would limit ourselves to wave functions that are \emph{even} under the interchange \(I \leftrightarrow II\) This could also mean that all information in region \(I\) is accurately copied into region \(II\), and this remains to be the case for the Fourier transforms (the Fourier transform of an even function is even as well.

This brings us very close to an ancient proposal by the author\,\cite{GtHambiguity.ref}, to identify the states in region \(II\) as the set of bra states associated to the kets in region \(I\). At first sight this still may generate problems, since energy reversal from \(I\) to \(II\) requires us to take real functions to be even but imaginary functions to be odd. This property however, would not be passed on correctly by the Fourier transforms. We suspect that this problem can be solved by restricting ourselves to \emph{real} wave functions only. 

Limiting ourselves to real wave functions only, now leads to serious questions  concerning the very foundations of quantum mechanics. The picture we obtained so-far is that there are three representations for the wave functions of the in- and out-particles: first, we have the original Schwarzschild metric, and all in- and out-going particles can be represented as wave functions depending on \(r,\ t,\ \tht,\) and \(\vv\). We subsequently write the same metric in terms of the tortoise coordinates \(x,\ y,\ \tht,\) and \(\vv\). Using the same wave functions, this gives us the quantum states in region \(I\) when \(x\) and \(y\) are positive. Finally, we have the negative values of \(x\) and \(y\). Again, the same wave functions now represent what goes on in region \(II\).

The equations for these wave functions in \((x,\,y)\) space will be as usual. The problems just mentioned above concern the time reversal in region \(II\). It is more than just an administrative difficulty. It seems that the concepts \emph{lowest energy state} and \emph{highest energy state} are interchanged there. If this is allowable then we can treat he entire system in terms of a single set of quantum field theoretical equations, where the technical difficulties in gluing the positive and negative values of \(x\) and \(y\) together must be more carefully addressed than we were able to do here.

\newsecl{On the Hawking temperature}{temperature.sec}

 An issue not yet addressed in detail is the value of Hawking's temperature,
 	\be kT_\Hawk=\frac 1{8\pi G M_\BH}\,. \eel{Hawktemp.eq}
 It is connected to the calculation of the Hawking entropy,
 	\be \bet=\frac 1{kT}=\frac{\pa S}{\pa E} = 8\pi G M\ .\eel{entropy.eq}
Now these values represent the contributions of the quantum states both in regions \(I\) and \(II\). If however, if region \(II\) is nothing but a quantum clone of region \(I\) then it should not contribute to the entropy at all, so that only half of the entropy is left:
	\be S=\half\cdot 4\pi G\,M^2 \ ; \qquad\bet=\half\,\bet_\Hawk\ , \qquad kT=2kT_\Hawk\ . \eel{double.eq}
This happens to be the temperature we anticipated in our treatment of Ref.\,\cite{GtHambiguity.ref} using different arguments: the picture of regions \(I\) and \(II\) together as representing bras and kets, turns our expressions into a density matrix. to arrive at expectation values one should not take the  absolute square of a density matrix element but instead, trace it with the relevant operators. Replacing the quadratic expressions by linear ones reduces the entropy by a factor 2.

Now both arguments mentioned above, to indicate that the value of the temperature is affected by a factor 2, seem to be mostly hand-waving. Author warns Reader against believing that \(N\) hand-waving arguments may add up to something more believable than a single one. But we have a more powerful argument, using the periodicity of the evolution operator in the Euclidean time direction. This was the calculation first employed by Hawking\,\cite{Hawking.ref} and others. 

Consider  an outside observer \(A\), and an observer \(B\) in the effectively flat spacetime at the origin of the Penrose diagram,  Figs.~\ref{Schw.fig}b and \ref{Penrose19.fig}.  If the outside observer \(A\) performs a time translation by a constant amount, 
	\be U(\del t)=\ex{-i \del t \,H}. \ee
 The tortoise variable \(x\) in Eqs. \eqn{tortoise.eq} is multiplied by \(\ex{\,\del t\,/\,4GM}\) and \(y\) is divided by the same amount. For observer \(B\) that is a Lorentz transformation at the origin. Now take the same transformation but replace \(\del t\) by an imaginary number \(-i\bet\). This is a rotation:
	\be x\ra x\, \ex{-i\bet /4GM}\ ,\qquad y\ra y\,\ex{i\bet /4GM} \ . \ee
Writing \be x=\rr\,\ex{i\vv}\ ,\qquad y=\rr\,\ex{-i\vv}\ ,\ee
one gets a real, Euclidean spacetime in polar coordinates (by using the coordinates \(x+y\) and \(-i(x-y)\)). The standard calculation, computing the free energy \(F\) from \(\Tr (\ex{-\bet H})\),  now assumes that only full rotations,
\(\vv\ra \vv+2\pi n\) lead spacetime back to its original orientation. The Green functions to be considered are therefore periodic with period \(2\pi\) in \(\vv\). This is how \(\bet\) takes the value \(8\pi GM\), The identification of the Euclidean periodicity \(\bet\) with \(1/kT\) is well-known 
 in condensed matter physics and in lattice theories for elementary particles, Thus one gets  Eq.~\eqn{entropy.eq} and \eqn{Hawktemp.eq} for the Hawking temperature.

Now in our theory, we have that region \(II\) has \(x\) and \(y\) replaced by \(-x\) and \(-y\). There, it  is a quantum clone of region \(I\). All our green functions must be the same in region \(II\) and in region \(I\) Thus, our Green functions must return to their original values when \(\vv\) is replaced by \(\vv+\pi\). This gives us the temperature \eqn{double.eq}.

 \newsecl{Conclusion}{conc} In some sense, our theory for black holes is extremely conservative.
 In our treatment there is no place for `entanglement' issues, and not even for attaching significant roles for the `Page time'\,\cite{Page.ref}.  Page   asks for explicit attention to very large time scales, where a significant fraction of the initial entropy is carried away by the radiated particles. But as soon as the radiated particles have moved further than a few Planck lengths away from the horizon crossing point, conventional laws of physics take over, entanglement or not. Any respectable theory for the production of these particles must include a chapter on how  these particles are expected to make this transition to free or almost free particle states in the outside world.

In fact, this seems to be easy enough. The total mass \(M\) of a black hole together with the surrounding particles is strictly constant. But only the particles that are still -- or already -- close to their respective horizons, contribute to the mass in practice. So, as soon as we see particles far enough away from the black hole, we should modify our description in terms of a black hole with mass \(M'\), slightly different from \(M\), by subtracting the contribution of the particles that are sufficiently far away, at any particular moment of time.

Note that, in our approximations, the masses of the radiated particles are small compared to that of the black hole itself, so that the time-dependent corrections, as described above, are infinitesimal in any case.
  I think we performed important ground work to formulate a precise framework for handling the physics of black holes. In ancient times when solitons, magnetic monopoles, field theoretical strings and membranes were regarded as novel predictions of gauge theories and related concepts were discovered so that their properties were calculated, no new physical principles had to be called upon. For black holes, including of course the rotating and charged versions, the situation is not at all that simple, and it could even be that the physical laws at small scales will have to be thoroughly revised in order to obtain a harmonic picture of what goes on. This would actually be welcome, since out present views of Planckian scale physics appear  to lead to paradoxes: how should one handle energies beyond that scale, and how can we reconcile that with quantum mechanics?
  
 The author's personal view is that both quantum mechanics and general coordinate invariance are aspects of matter and geometry that are calling for a more precise formulation of their foundations. It does not help to be `agnostic', as some physicists declared, or to think that `chaos' takes care of all contradictions. 
 
Particularly, our results on Hawking's temperature are novel (they were already mentioned in 1984\,\cite{GtHambiguity.ref} but hardly noticed by string theorists). Some further thinking strongly suggests that, actually, any attempt to identify the information residing in region \(II\) with what is already present in region \(I\), would reduce the Hawking entropy by a factor 2, and raise the temperature of the radiation accordingly.

Since conservation of information (in the sense of entropy in thermodynamics) is being considered more seriously these days, and in view of our considerably revised outcome \eqn{double.eq} for  temperature and entropy in black holes, we conclude that thorough revisions for quantum gravity itself may be opportune. Our proposals for real wave functions rather than complex ones, also suggests revisions for the foundation of quantum mechanics\,\cite{fast.ref}.

We thank N.~Sanchez, N.~Gaddam, F.~Feleppa, N. Groenenboom, and S.~Kumar for  discussions.

 \end{document}